\documentstyle[epsf]{elsart}
\begin{document}
\begin{frontmatter}
\noindent To appear in Proceedings of the Dynamics Days Asia Pacific 
Conference, 13-16 July, 1999, Hong Kong (Physica A, 2000).\\

\title{Modelling High-frequency Economic Time Series
}
\author{Lei-Han Tang$^a$ and Zhi-Feng Huang$^{a,b}$}
\address{$^a$Department of Physics and Center for Nonlinear Studies\\
Hong Kong Baptist University, Kowloon Tong, Hong Kong\\
$^b$
Center for Advanced Study,
Tsinghua University,
Beijing 100084,
P. R. China
}

\begin{abstract}
The minute-by-minute move of the Hang Seng Index (HSI) data
over a four-year period is analysed and shown to possess
similar statistical features as those of other markets.
Based on a mathematical theorem 
[S. B. Pope and E. S. C. Ching, Phys. Fluids A {\bf 5}, 1529 (1993)], 
we derive an analytic form for 
the probability distribution function (PDF) of index moves 
from fitted functional forms of certain conditional averages 
of the time series. Furthermore, following a recent work by
Stolovitzky and Ching, we show that the observed PDF
can be reproduced by a Langevin process with a move-dependent noise
amplitude. The form of the Langevin equation can be determined directly
from the market data.
\end{abstract}
\end{frontmatter}

The availability of high-frequency economic time series,
with a sampling rate of every few seconds, has generated
a great deal of theoretical interest in the econometrics and the
econophysics community\cite{net,ms95,ms_book,bp_book}. 
Attempts have been made to devise models which
produce time series with similar statistical characteristics
as those of real markets. Many of these studies are based on
variants of the Autogressive Conditional Heteroskedasticity (ARCH)
process first introduced by Engle\cite{engle} to analyze the quarterly 
consumer price index in the UK over the period 1958 to 1977,
and the generalised ARCH (GARCH) process which offers a more
flexible description of the volatility memory effect 
(i.e., lag structure)\cite{boll}. 
The nonlinearity in the regression models makes it possible to
generate probability distribution functions (PDF) with 
fat tails, a characteristic of financial data first noted by 
Mandelbrot\cite{m62}.
However, since all these processes are discrete
in time, an immediate question to ask is whether the quality of the 
modelling depends on the time unit chosen and if there is
a time scale which is the most natural of all. 
Indeed, when the time step is not chosen properly, one
has to either introduce many terms in the regression expression
[the GARCH($p,q$) model] to take into account memory effects,
or to miss some of the important short-time statistics.

An alternative approach, which partially circumvents the above
difficulty, is to model the market price move as a continuous time process.
Continuous time stochastic processes are quite familiar to physicists,
ranging from simple Brownian motion to the fully-developed turbulence. 
In fact, the high-frequency market price movements have much in
common with the velocity or temperature time series in 
turbulent flows\cite{gbptd,ms97,cast,turb},
an analogy we exploit in this paper.
To put this statement on more quantitative terms, let us first
summarise two salient statistical features which seem to be 
universally true for all major stock indices\cite{bp_book}.

\begin{figure}
\epsfxsize=10truecm
\epsfbox{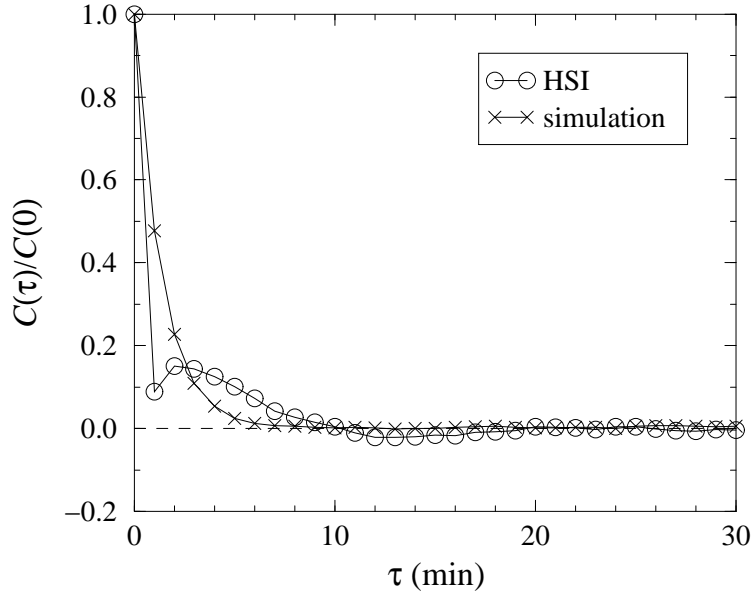}
\caption{Normalised linear two-point correlation function of the 
minute-by-minute HSI moves over the four-year period 1994 to 1997. 
Note the weak oscillations of the correlation function, indicating
a slightly under-damped behaviour. Also shown is the same 
correlation function calculated from the simulation using a
Langevin equation.}
\label{fig1}
\end{figure}

(i) {\it Short linear correlation of price moves} ---
For a given stock index $S(t)$, one may define the price move
over a fixed time interval $\delta$ (say one minute),
\begin{equation}
x(t)=S(t)-S(t-\delta).
\label{move}
\end{equation}
It has been shown that the ``linear correlation''
\begin{equation}
C(\tau)=\langle x(t+\tau)x(t)\rangle
\label{linear-corr}
\end{equation}
decays to zero very rapidly, on the order of ten minutes.
We have analysed the minute-by-minute Hang Seng Index (HSI) data
collected over a four year period January 1994 --- December, 1997.
Figure 1 shows the linear correlation function $C(\tau)$ 
with $\delta=1$ min. It is seen that $C(\tau)$ becomes nearly
zero after a period of ten minutes or so. The decay is however
not completely monotone, suggesting that the market is slightly
under-damped.

(ii) {\it Nongaussian distribution of price moves with fat tails} ---
The fat tails of the PDF $P(x)$ of
stock price moves are well-known and have also been observed for 
the movement of foreign currency exchange rates. Mandelbrot has
observed that $P(x)$ often decays as a power-law function of $|x|$,
and hence, combining with (i), the stock index can be considered
as a realisation of L\'evy walk\cite{m62}. 
From the analysis of the high-frequency
S\&P 500 data, Mantegna and Stanley showed that a truncated
L\'evy distribution offers a better description of the PDF\cite{ms95}.
Figure 2 shows the PDF for the HSI minute-by-minute move
data $x(t)$ (open circles) on a semi-log scale, 
collected over the same period as in Fig. 1. It is seen that
the decay at large $|x|$ can be well-described by a simple exponential
function, as observed previously in Ref. \cite{ms95}. 
For small $|x|$, a different behaviour is seen.
A noticeable feature is the cusp-like singularity at $x=0$,
which is so far unexplained.

\begin{figure}
\epsfxsize=10truecm
\epsfbox{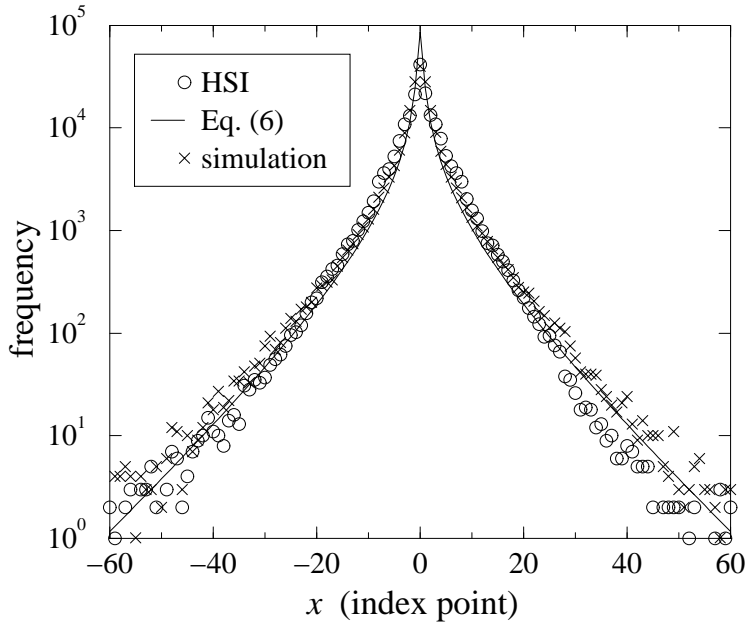}
\caption{The (unnormalised) probability distribution function (PDF) of the
minute-by-minute HSI move ($N=201704$ events). 
Data in the first twenty minutes
of each morning trading session are discarded.
Also shown are the PDF calculated from Eq. (\ref{hsi-pdf}) by fitting
the conditional averages (solid line), and the PDF of the
time series from the simulation (crosses).}
\label{fig2}
\end{figure}

The peculiar form of the PDF as seen in Fig. 2 has in fact been observed
in a physical context\cite{cast}. In analysing the temperature
time series of thermal convection in the ``hard-turbulence'' regime,
Pope and Ching\cite{turb} considered the following conditional 
averages for a twice-differentiable time series $x(t)$,
\begin{equation}
r(x)={\langle \ddot{x}|x\rangle\over \langle\dot{x}^2\rangle},\qquad
q(x)={\langle \dot{x}^2|x\rangle\over\langle\dot{x}^2\rangle}.
\label{r_xq_x}
\end{equation}
Here $\langle\cdot|x\rangle$ denotes the average of a given
quantity over those data points in the time series where $x(t)=x$.
From the stationarity of the PDF, they proved that
the PDF and the conditional averages $r(x)$ and $q(x)$ are related
through the following equation,
\begin{equation}
P(x)={C\over q(x)}\exp\Bigl[\int_0^x{r(x')\over q(x')}dx'\Bigr].
\label{pdf}
\end{equation}
Using the turbulent temperature time series data as input,
they showed that $r(x)$ is generally linear in $x$ with a negative
coefficient. In the soft turbulence regime, $q(x)$ is nearly
constant. From Eq. (\ref{pdf}), the resulting PDF is gaussian as
observed. On the other hand, in the hard turbulence regime,
$q(x)$ increases with increasing $|x|$, giving rise to fat tails
in the PDF.

\begin{figure}
\epsfxsize=13truecm
\epsfbox{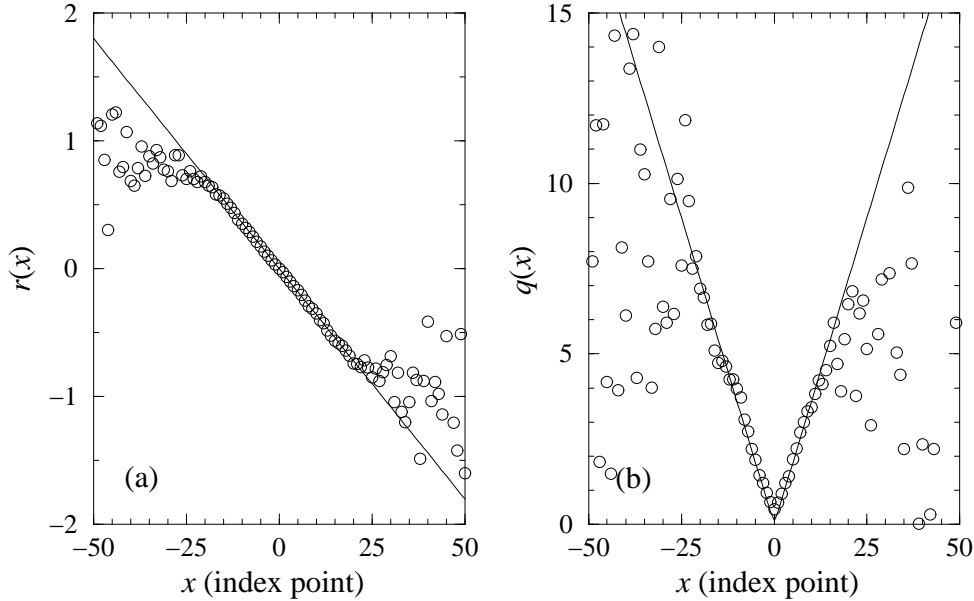}
\caption{
(a) Conditional average $r(x)$ (circles) and the linear fit
(solid line). (b) Conditional average $q(x)$ (circles) and a nearly linear
fit (solid line). 
}
\label{fig3}
\end{figure}

Figure 3 shows $r(x)$ and $q(x)$ computed using the HSI move time series.
Indeed, the shape of these two functions are very much like the temperature
data in Ref. \cite{turb}, although there are small differences. 
The data can be fitted to the following functional forms,
\begin{equation}
r(x)=-Rx,\qquad q(x)=Q(x^2+a^2)^{1/2},
\label{fit}
\end{equation}
where $R=0.036$ and $Q=0.36$. The round-off parameter $a$ can not
be determined precisely because of the discreteness of the index data.
Instead, we fix its value from the normalisation condition.
Substituting Eq. (\ref{fit})
into Eq. (\ref{pdf}), and carrying out the integration, we obtain,
\begin{equation}
P(x)={C'\over (x^2+a^2)^{1/2}}\exp\Bigl[-\alpha(x^2+a^2)^{1/2}\Bigr],
\label{hsi-pdf}
\end{equation}
where $\alpha=R/Q=0.1$ and $C'^{-1}=2K_0(\alpha a)$, with
$K_0(u)$ being a modified Bessel function of the second kind. 
The solid line in Fig. 2 is produced by taking $a^2=0.1$.
As can be seen, the agreement between the original 
data and the fitted form (\ref{hsi-pdf}) is rather satisfactory. 

It is worth noting that the functional form (\ref{hsi-pdf}) 
we propose is quite different from those of earlier studies\cite{note}. 
The behaviour of the PDF
around $x=0$ is controlled by the parameter $a$. A sharp peak
is produced when $a$ becomes very small. In this respect,
$a$ serves the purpose of a cut-off related to the discreteness
of the underlying asset price. For large $|x|$,
$P(x)$ crosses over to simple exponential decay. There is
however no scale invariance as previously suggested.

The more challenging task is to devise a dynamic equation that
generates a time series with the same conditional averages as
those of the market data. This issue was considered recently
by Stolovitzky and Ching\cite{stol}. They studied a one-dimensional
Langevin process defined by the following stochastic differential equation,
\begin{equation}
m{d^2 x\over dt^2}+\gamma{dx\over dt}=F(x)+[2\gamma k_BT(x)]^{1/2}\xi(t),
\label{langevin}
\end{equation}
where $\xi(t)$ is a gaussian white noise with $\langle \xi(t)\rangle=0$
and $\langle \xi(t)\xi(t')\rangle=\delta(t-t')$.
The main difference of (\ref{langevin}) from the usual Brownian
process is an $x$-dependent temperature (or noise amplitude)
$T(x)$. In the over-damped limit $\gamma\rightarrow\infty$, they showed
that the conditional averages are given by,
\begin{equation}
\langle\ddot{x}|x\rangle=F(x)/m,\qquad
\langle\dot{x}^2|x\rangle=k_BT(x)/m.
\label{large-gamma}
\end{equation}
Combining with Eq. (\ref{pdf}), Stolovitzky and Ching showed
that the PDF in this limit is given by a generalised Boltzmann form,
\begin{equation}
P(x)={C\over T(x)}\exp\Bigl[\int_0^x {F(x')\over k_BT(x')}dx'\Bigr].
\label{Boltzmann}
\end{equation}

The significance of the above result is as follows.
Assuming that a given time series is generated by a Langevin 
process (\ref{langevin}), one can then determine the
effective force $F(x)$ and the effective temperature $T(x)$ uniquely
by computing the conditional averages from the data,
apart from an overall time constant. The PDF of the Langevin time series
is identical to the PDF of the original data by construction.

The $\gamma\rightarrow\infty$ limit is quite suitable for
the analysis of the HSI data, as we have seen from the two-point
correlation function $C(\tau)$ that any memory effect about the
direction of the move decays to zero rapidly.
It is then suggestive to drop the inertia (i.e. mass) term in
Eq. (\ref{langevin}) altogether. Performing the scaling
$t\rightarrow \gamma t$ and setting $k_B=1$, we can cast
Eq. (\ref{langevin}) in the form,
\begin{equation}
{dx\over dt}=F(x)+[2T(x)]^{1/2}\xi(t).
\label{langevin1}
\end{equation}
The effective force and the effective temperature are related
to the conditional averages through Eq. (\ref{large-gamma}).
The parameter $m$ can be chosen to include a short-time relaxation
effect as seen in Fig. 1\cite{note2}. 

We have simulated the Langevin equation (\ref{langevin1}) using
an Euler integration scheme with $\Delta t=0.1$ min. The form
of the functions $F(x)$ and $T(x)$ are determined from the conditional
averages (\ref{fit}). The PDF of the simulated
minute-by-minute moves, with the same number of events as the
original data, is shown in Fig. 2 (crosses). In Fig. 4 we plot
both the HSI data (daily close) and the simulated index data with 
an artificial annual yield of $10\%$. The gross features of the two data 
sets seem to be similar to the naked eye, though in the simulated data set
there is no daily and weekly breaks.

\begin{figure}
\epsfxsize=10truecm
\epsfbox{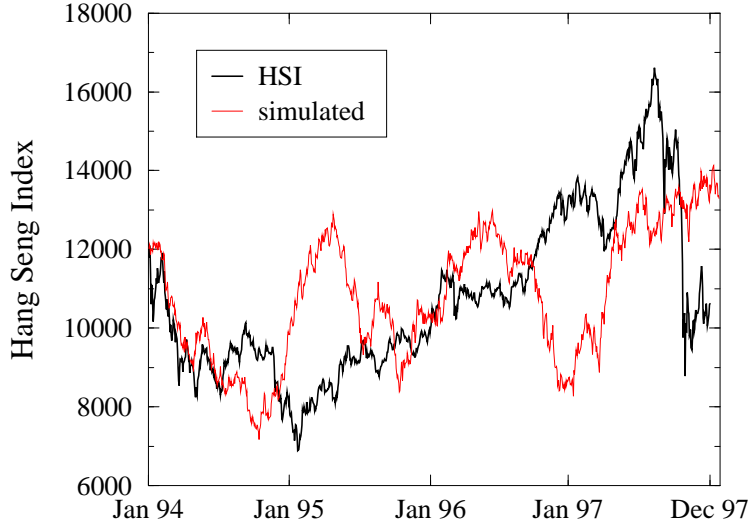}
\caption{
Daily close of the HSI over a four year period (heavy line)
and the simulated index data (dotted line). 
}
\label{fig4}
\end{figure}

One important feature which is missing in the Langevin equation
(\ref{langevin}) is the long-term volatility correlations which
may last from a few days to several weeks or longer. The simulated
time series has essentially the same relaxation time, on the order
of a few minutes in our case, for the linear move correlation 
and for the volatility correlation. It is however possible to introduce
volatility persistence by hand into the simulation. The effect of
a nonstationary volatility on the PDF of the time series is a subject
under current investigation.

{\bf Acknowledgements:} 
We would like to thank Prof. Lam Kin at the Department of Finance and
Decision Sciences, HK Baptist University for providing the 
HSI data, and Dr. Emily Ching for sending us Ref. \cite{stol}
prior to publication. The work is supported in part by the Hong Kong
Baptist University under grant FRG/97-98/II-78.


\begin{thebibliography}{9}
\bibitem{note} In Ref. \cite{bp_book} several classes of distributions
were discussed. The closest to ours is probably the hyperbolic
distribution. However, without the denominator, the PDF
does not possess a sharp peak at $x=0$ which is one of the
defining properties of high-frequency data.
\bibitem{note2} The slightly under-damped behaviour seen in Fig. 1
can be simulated by including both the mass term and the friction
term. Details of our study will be reported elsewhere.
\end{thebibliography}
\end{document}